\documentclass[12pt]{article}

\usepackage{graphicx}
\usepackage{dcolumn}
\usepackage{bm}
\usepackage{epsfig}

\begin{document}

\title{Aging and fluctuation-dissipation ratio in a nonequilibrium 
$q$-state lattice model}

\author{M. O. Hase, T. Tom\'e and M. J. de Oliveira \\
Instituto de F\'{\i}sica, Universidade de S\~{a}o Paulo, \\
Caixa Postal 66318, \\
05314-970 S\~{a}o Paulo, S\~{a}o Paulo, Brazil}
\date{}

\maketitle

\begin{abstract}

A generalized version of the nonequilibrium linear Glauber model
with $q$ states in $d$ dimensions is introduced and analyzed.
The model is fully symmetric, its dynamics being invariant under
all permutations of the $q$ states.
Exact expressions for the two-time autocorrelation and response
functions on a $d$-dimensional lattice are obtained. In the stationary
regime, the fluctuation-dissipation theorem holds, while in the transient
the aging is observed with the fluctuation-dissipation ratio leading to
the value predicted for the linear Glauber model.


\end{abstract}

\section{Introduction}
\label{introduction}

The model introduced originally by Glauber \cite{G63} 
is defined on a one dimensional lattice 
where a one-spin-flip Markovian stochastic process takes place. 
It simulates the dynamics of a ferromagnetic Ising chain with 
first-neighbor interaction, and the stationary state is described 
by the Gibbs measure associated with the Ising Hamiltonian as a 
consequence of obeying detailed balance. In stationary regime at 
non-zero temperatures, the two-time functions like the autocorrelation 
and response function, are time-translationally invariant and connected 
through the fluctuation-dissipation theorem; furthermore, the model 
displays a transient where aging is observed \cite{GL00a}. At zero 
temperature, when the system becomes critical, the fluctuation-dissipation 
ratio (which is related to the effective temperature of the system\cite{CKP97}) assumes a nontrivial value $X_{\infty}=1/2$.

When the model is extended to higher dimensions, and called
Glauber model, no analytical 
solution is available due to the nonlinear structure of the transition rate. 
Nevertheless, the linearized version of the Glauber model, proposed a 
few years ago \cite{dO03}, can be treated by analytical tools in any 
dimension. The linear Glauber model can be seen as the voter model 
with noise, which displays a disordered (paramagnetic) phase only; 
in the absence of noise, however, the system becomes critical.
The dynamics of linear Glauber model was investigated by some of the 
authors in a previous paper \cite{HSTdO06}. This model is microscopically 
irreversible, that is, it does not obey detailed balance in the 
stationary state, for dimension $d>1$. Moreover, it has both the 
stationary and aging regimes, with a nontrivial fluctuation-dissipation 
ratio $X_{\infty}=1/2$ in the later regime as in the usual one-dimensional 
Glauber case.

The fluctuation-dissipation relation was usually conceived for systems 
that obey detailed balance \cite{CW51, K57}, and has been generalized to 
include non-stationary regimes by the introduction of an effective 
temperature measuring the violation of fluctuation-dissipation 
theorem (\cite{CR03}, and references therein). Many works have confirmed 
this phenomenon for several models 
\cite{CD95, GL00b, GL01, CST01, HS06, WC08, WC09}. 
Recent progress has suggested that it can also be invoked 
for nonequilibrium models \cite{HSTdO06,dO07}, 
which does not have an associated known Hamiltonian (see also this
issue in the context of kinetically constrained models\cite{RS03, LMSBG07}). 
In \cite{dO07}, the fluctuation-dissipation relations were analyzed in 
a general class of models that exhibit up-down symmetry which does not 
obey detailed balance.

The above mentioned works are closely related to the question of 
universality in out-of-equilibrium processes \cite{C04}. In equilibrium 
statistical mechanics, it is widely known that the critical behaviour 
of a system is governed by the fixed point of the renormalization 
transformation, and it turns out that only a few characteristics of 
the model are relevant to determine its universality class.

In out-of-equilibrium dynamics, some of the problems related to 
universality may be 
addressed by the generalized version of the fluctuation-dissipation theorem
\begin{eqnarray}
R(t,t^{\prime})=X(t,t^{\prime})\frac{\partial}{\partial t^{\prime}}
C(t,t^{\prime}),
\label{fdt}
\end{eqnarray}
where $R(t,t^{\prime})$ and $C(t,t^{\prime})$ are the response function 
and autocorrelation, respectively (see \cite{LCSZ08} for some recent results). The typical experimental situation 
under consideration is a quench from a completely disordered state,
which in reversible systems corresponds to a high temperature state,
to the critical point. 
The usual fluctuation-dissipation relation is verified when the 
fluctuation-dissipation ratio $X(t,t^{\prime})$ equals the unity.
It was conjectured that $X(t,t^{\prime})$ would depend functionally on 
$C(t,t^{\prime})$ only \cite{CK94}, but renormalization group analysis 
\cite{CG02} and numerical calculations \cite{C03} indicated that the 
fluctuation-dissipation ratio is a function of $t/t^{\prime}$. 
Furthermore, scaling arguments were casted to propose the asymptotic 
behaviour for the autocorrelation and response function \cite{GL00b} 
at the critical temperature. This result suggested that the quantity
\begin{eqnarray}
X_{\infty}=\lim_{t^{\prime}\to\infty}\lim_{t\to\infty}X(t,t^{\prime})
\label{Xinf}
\end{eqnarray}
is universal due to its dependence to dynamical exponents and the 
ratio of autocorrelation and response amplitudes, which are 
conjectured to be universal \cite{GL00a}.

In this paper, we introduce and analyze a nonequilibrium lattice model, 
which is a generalization of the nonequilibrium linear Glauber model
to more than two states, which we call linear $q$-state model. 
The model can be understood as
the linearized version of the dynamics associated to 
the equilibrium $q$-state Potts model. The dynamics of the
model, as is the case of any dynamics of the Potts model, 
is invariant under the permutation of any two states.
The model can also be understood as a $q$-state voter model with noise.
In this interpretation, a group of individuals are called
to vote in one of $q$ candidates. A voter changes his 
opinion by choosing randomly a neighbor individual and adopting the
neighbor's opinion with probability $\mu$ and remaining with
his opinion with probability $1-\mu$, the noise. 
Without noise, it reduces to the ordinary 
$q$-state voter model \cite{HG98}. 
Similarly to the linear Glauber model, the present
linear $q$-state model displays a paramagnetic phase whenever 
$0<\mu<1$ and becomes critical at $\mu=1$.

The analysis of such model has two main aims. 
First, it addresses the question raised in a previous 
result \cite{HSTdO06} about the dynamical phenomena of aging and violation 
of fluctuation-dissipation for a class of systems that does not obey 
detailed balance -- recall that these problems were usually studied 
through models that are described by a Hamiltonian. Finally, there is 
an additional interest in considering a model with a more general 
symmetry in order to verify its influence on the (possibly) universal 
quantity cited above, since in equilibrium statistical physics, symmetry 
plays a major role in the critical behaviour.

The equilibrium Potts model with $q$ states has been used to describe experimentally systems that display a number of identical states or structures at low temperatures \cite{W82} such as the adsorption of noble gases on graphite \cite{W82}. It has also been used to describe biological cell sorting \cite{GG92}. The nonequilibrium model with many equivalent states such as the one studied here may be relevant in the description of systems where microscopic reversibility is not ensured like some biological phenomena\cite{GG92}.

The layout of this paper is as follows. In section \ref{qstate}, 
the nonequilibrium linear $q$-state 
model is defined and many one-time functions are determined analytically
The two-time functions are calculated in section \ref{CR}, 
where the fluctuation-dissipation 
relations are carefully examined, and some dynamical exponents are 
calculated in section \ref{dynamicalexponents}.
The summary of the main results and its 
discussions are found in the last section.

\section{The linear $q$-state model}
\label{qstate}

Consider a $d$-dimensional hypercubic lattice 
with $N=L^d$ sites and periodic boundary conditions.
To each site $i$ there is a spin variable
$\sigma_i$ that takes the values $0,1,\ldots,q-1$.
The time evolution is governed by a one-site dynamics
in which the state of a given site $i$ changes from 
$\sigma_i$ to $\sigma_i^\prime = \sigma_i + \alpha$ modulo $q$,
where $\alpha$ is one of the $q$ states,
and the states of the other sites remain unchanged.  
The possible transitions are then the ones in which the state 
$\sigma = (\sigma_1,\sigma_2,\ldots,\sigma_i,\ldots,\sigma_N)$
changes to the state 
$\sigma^{i,\alpha}=(\sigma_1,\sigma_2,\ldots,\sigma_i^\prime,\ldots,\sigma_N)$
where $\sigma_i^\prime = \sigma_i + \alpha$ modulo $q$.
The corresponding transition rate is denoted by $w_i^\alpha(\sigma)$
and, for the nonequilibrium linear $q$-state model, 
is defined by
\begin{eqnarray}
w_{i}^\alpha(\sigma)=\frac{1-\mu}{q}+\frac{\mu}{2d}\sum_{\delta}
\delta(\sigma_{i}+\alpha,\sigma_{i+\delta}),
\label{wa}
\end{eqnarray}
where the summation is over the nearest neighbors 
and $\delta(x,y)$ is the Kronecker delta, 
which equals $1$ if $x=y$ and $0$ otherwise
and the parameter $\mu$ takes values in the interval
$0<\mu\leq1$.
The time evolution
of the probability $P(\sigma,t)$ of finding the system
at state $\sigma$ at time $t$ is governed by the
master equation
\begin{equation}
\frac{d}{dt}P(\sigma,t) = \sum_i\sum_{\alpha}
\Big[ w_{i}^{\alpha}(\sigma^{i,-\alpha}) P(\sigma^{i,-\alpha},t)
- w_{i}^{\alpha}(\sigma)P(\sigma,t) \Big],
\label{meq}
\end{equation}
where the summation in $\alpha$ extends over the $q$ states.

The probability of a spin at site $j$ be at state, say $1$, is given 
by $\langle\delta(\sigma_{j},1)\rangle$. Throughout this paper, the notation
\begin{eqnarray}
\langle A(\sigma)\rangle = \sum_{\sigma}A(\sigma)P(\sigma,t)
\end{eqnarray}
will denote the average over spin configurations
of the state function $A(\sigma)$.
The equation of motion for 
$\langle\delta(\sigma_{j},1)\rangle$ can be written from the master 
equation (\ref{meq}) as
\begin{eqnarray}
\frac{d}{dt}\langle\delta(\sigma_{j},1)\rangle 
= \frac{1-\mu}{q}-\langle\delta(\sigma_{j},1)\rangle
+\frac{\mu}{2d}\sum_{\delta}\langle\delta(\sigma_{j+\delta},1)\rangle.
\label{d<d>}
\end{eqnarray}
It is also possible to describe the time evolution of 
$\langle\delta(\sigma_{j},1)\delta(\sigma_{k},1)\rangle$,
\begin{eqnarray}
\nonumber \frac{d}{dt}\langle\delta(\sigma_{j},1)\delta(\sigma_{k},1)\rangle & 
= & \left(\frac{1-\mu}{q}\right)\Big[ \langle\delta(\sigma_{j},1)\rangle
+\langle\delta(\sigma_{k},1)\rangle \Big] - 2\langle\delta(\sigma_{j},1)
\delta(\sigma_{k},1)\rangle + \\
 & & + \frac{\mu}{2d}\sum_{\delta}\Big[ \langle\delta(\sigma_{j+\delta},1)
\delta(\sigma_{k},1)\rangle+\langle\delta(\sigma_{k+\delta},1)
\delta(\sigma_{j},1)\rangle \Big], \quad j\neq k,
\label{d<dd>}
\end{eqnarray}
which is closely related to the pair correlation. In order to 
recover the results obtained by \cite{dO03} and \cite{HSTdO06}
for the linear Glauber model,
it is necessary to connect $\delta(\sigma_{j},1)$ 
to an Ising spin variable $s_j$, that takes the values $-1$ or $+1$, 
through the relation
\begin{eqnarray}
s_{j} = 2\delta(\sigma_{j},1)-1.
\label{q=2}
\end{eqnarray}

\begin{figure}[tbp]
\centering
\epsfig{file = 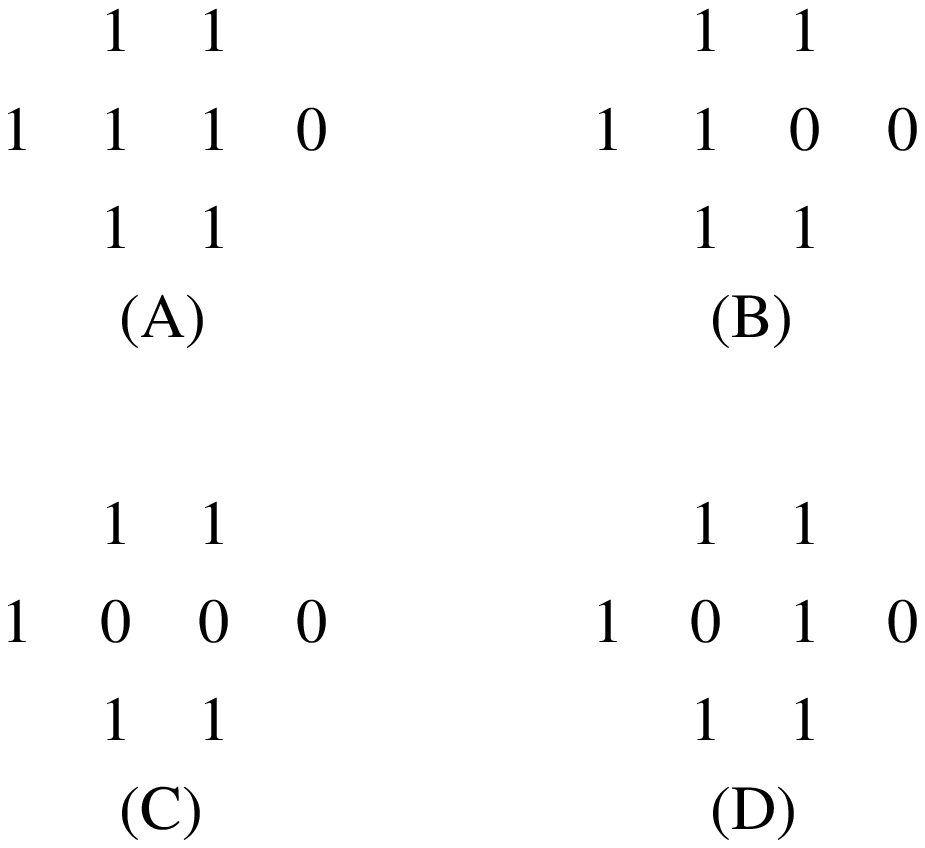, scale = 0.7}
\caption{A possible irreversible sequence for the bidimensional $q$-state model ($q\geq 2$).}
\label{irreversible}
\end{figure}

As a last remark on the model, its irreversible property for $d\geq 2$ will be discussed. Consider, for instance, the four states shown in figure \ref{irreversible} on a square lattice ($d=2$). Suppose that the system follows the sequence of states $A$, $B$, $C$ and $D$ and returns to the initial state $A$. If the interval between two successive states is $\Delta t$, the probability of occurrence of the sequence $A\rightarrow B\rightarrow C\rightarrow D\rightarrow A$ can be calculated through the transition rate $w_{i}$ (which is $w_{i}^{\alpha=0}(\sigma)$) as
\begin{eqnarray}
\nonumber P(A\rightarrow B\rightarrow C\rightarrow D\rightarrow A) & = & P(A|D)P(D|C)P(C|B)P(B|A)P(A) \\
 & = & \left(\frac{1-\mu}{q}+3\frac{\mu}{2d}\right)^{2}\left(\frac{1-\mu}{q}+2\frac{\mu}{2d}\right)\left(\frac{1-\mu}{q}\right)\,.
\label{path1}
\end{eqnarray}
This result is not necessarily equal to the probability of observing the reveresed sequence $A\rightarrow D\rightarrow C\rightarrow B\rightarrow A$, which is
\begin{eqnarray}
\nonumber P(A\rightarrow D\rightarrow C\rightarrow B\rightarrow A) & = & P(A|B)P(B|C)P(C|D)P(D|A)P(A) \\
 & = & \left(\frac{1-\mu}{q}+4\frac{\mu}{2d}\right)\left(\frac{1-\mu}{q}+2\frac{\mu}{2d}\right)\left(\frac{1-\mu}{q}+1\frac{\mu}{2d}\right)^{2}\,.
\label{path2}
\end{eqnarray}
The existence of a sequence of states that is not reversible implies that the system is irreversible. A generalization of this result to higher dimensions is obtained by, for instance, filling the sites created by the introduction of more dimensions with spins $\tilde \sigma$ where $\tilde\sigma\neq 0,1$ (this is the situation when $q>2$; the case $q=2$ was already discussed in \cite{HSTdO06}). The argument above can also be invoked to show that the model is reversible in the one-dimensional case.

\subsection{Site magnetization}
\label{site_m}

The definition of the site magnetization will be guided by some constraints. 
At a fully ordered state, where the spin is at state, say $\sigma_{j}=1$, one 
should have $\langle\delta(\sigma_{j},1)\rangle=1$, while at disordered 
state, where the spin is at any one of the $q$ state with equal probability, 
the condition $\langle\delta(\sigma_{j},1)\rangle=1/q$ should be satisfied. 
This leads to a natural definition of an ordered parameter 
$\sum_{j}m_{j}/N$, where the site magnetization is defined by
\begin{eqnarray}
m_{j}(t)=\frac{q}{q-1}\langle\delta(\sigma_{j},1)\rangle-\frac{1}{q-1}.
\label{m}
\end{eqnarray}
The above definition leads to
\begin{eqnarray}
\frac{d}{dt}m_{j}(t)=-m_{j}(t)+\frac{\mu}{2d}\sum_{\delta}m_{j+\delta}(t),
\label{dmdt}
\end{eqnarray}
which is the time evolution of the site magnetization.

The equation (\ref{dmdt}) can be solved by the usual methods by 
introducing, for instance, the Fourier transform
\begin{eqnarray}
m_{p}^{F}(t)
=\sum_{j}m_{j}(t)e^{-ijp}, 
\end{eqnarray}
and its inverse
\begin{eqnarray}
m_{j}(t) = \frac{1}{N}\sum_{p} m_{p}^{F}(t)e^{ijp},
\end{eqnarray}
where the summation in $p$ is over the sites of the
the first Brillouin zone, in which each component
of the vector $p$ takes values inside the interval
between $-\pi$ and $\pi$.
The solution of the differential equation (\ref{dmdt}) is then
\begin{eqnarray}
m_{j}(t)=\sum_{\ell}\Gamma_{j-\ell}(t-t^{\prime})
m_{\ell}(t^{\prime}),
\label{mt}
\end{eqnarray}
where
\begin{eqnarray}
\Gamma_{j}(t)=\frac{1}{N}\sum_{p}e^{ijp-f_{\mu}(p)t},
\end{eqnarray}
and
\begin{eqnarray}
f_{\mu}(p)=1-\frac{\mu}{d}\sum_{i=1}^{d}\cos p_{i},
\label{Gamma}
\end{eqnarray}
for a $d$-dimensional hypercubic lattice with coordination number $z=2d$. 
For a homogeneous initial condition $m_{j}(t^{\prime})=m^{0}$ for any 
$j$, it is straightforward that the site magnetization is constant 
($m_{j}(t)=m^{0}$) for $\mu=1$. On the other hand, the condition 
$\mu\neq 1$ implies 
$m_{j}(t)=m^{0}e^{-\left(1-\mu\right)\left(t-t^{\prime}\right)}$, 
which means that the magnetization decays to zero for sufficiently long time. 
The time correlation length $\overline\tau$, defined by 
$m=m^{0}e^{-t/\overline\tau}$, diverges as 
$\overline\tau\sim(1-\mu)^{-\nu_{\parallel}}$, 
from which $\nu_{\parallel}=1$. Moreover, the static magnetization 
$M=\lim_{t\to\infty}\sum_{j}m_{j}(t)/N$ is always zero for 
$\mu\neq 1$ and is a non-zero constant (if $m^{0}\neq 0$) 
at criticality ($\mu=1$); this jump in the magnetization implies $\beta=0$.

\subsection{Pair correlation}
\label{pair}

The definition of pair correlation $q_{j,k}(t)$ for spins at $j$ 
and $k$ will be gauged to obey the following requirements:
(i) $q_{j,j}(t)=1$ for any $j$, 
(ii) $q_{j,k}(t)=0$, $j\neq k$, for the paramagnetic state
(iii) $q_{j,k}(t)=1$, $j\neq k$, for the ordered state.
These conditions leads to a natural definition for the pair correlation, 
which is
\begin{eqnarray}
q_{j,k}(t)=\frac{q^{2}}{q-1}\langle\delta(\sigma_{j},1)\delta(\sigma_{k},1)
\rangle - \frac{q}{2}\Big[ m_{j}(t)+m_{k}(t) \Big] -\frac{1}{q-1},
\label{q}
\end{eqnarray}
and is the only one where the pair correlation $q_{j,k}(t)$ is linear to 
$\langle\delta(\sigma_{j},1)\delta(\sigma_{k},1)\rangle$. Its time evolution,
\begin{eqnarray}
\nonumber 
\frac{d}{dt}q_{j,k}(t) & = & -2q_{j,k}(t)
+\frac{\mu}{2d}\sum_{\delta}\Big[q_{j+\delta,k}(t)
+q_{k+\delta,j}(t)\Big] \\
& & - \frac{1}{2}\left(1-\mu\right)
\left(q-2\right)\Big[ m_{j}(t)+m_{k}(t) \Big], \qquad j\neq k,
\label{pre_dqdt0}
\end{eqnarray}
is obtained from the master equation (\ref{meq})and 
from equation (\ref{d<dd>}).

From now on, it will be assumed that the pair correlation $q_{j,k}(t)$ 
depends on sites $j$ and $k$ through their difference $r=j-k$ only 
(note that $q_{r}(t)=q_{-r}(t)$). Moreover, the system will be assumed 
to be in a random initial state (see \cite{GSPR05}, \cite{AS06} and \cite{CG07} for other possibilities), such that the evolution process
can be understood as a quench from $\mu=1/q$ to a $\mu\neq1/q$,
which implies $m_{j}(t)=0$ for any $t$. This condition leads to
\begin{eqnarray}
\frac{d}{dt}q_{r}(t) & = & -2q_{r}(t)
+\frac{\mu}{d}\sum_{\delta}q_{r+\delta}(t), \qquad r\neq 0.
\label{dqdt0}
\end{eqnarray}
The above equation, which is valid for $r\neq 0$ only, should be modified to comprise the case $r=0$, 
for which $q_0(t)=1$. By using a previously introduced method 
\cite{dO03}, we write the equation (\ref{dqdt0}) in the form
\begin{eqnarray}
\frac{d}{dt}q_{r}(t)=-2q_{r}(t)+\frac{\mu}{d}\sum_{\delta}q_{r+\delta}(t)
+b(t)\delta_{r,0},
\label{dqdt}
\end{eqnarray}
which is now valid for any $r$, including the case $r=0$,
provided $b(t)$ be chosen to ensure $q_{0}(t)=1$ (note that if $r\neq 0$, then equation (\ref{dqdt}) recovers (\ref{dqdt0})). Formally, this means that the function $b(t)$ should satisfy
\begin{eqnarray}
b(t) = \frac{d}{dt}q_{0}(t) + 2q_{0}(t) - \frac{\mu}{d}\sum_{\delta}q_{\delta}(t) = 2 - \frac{\mu}{d}\sum_{\delta}q_{\delta}(t)\,.
\label{br0}
\end{eqnarray}

Since the system was assumed to be in a completed disordered 
initial condition, then $q_{r}(0)=\delta_{r,0}$ and the equation 
of motion for the pair correlation (\ref{dqdt}) can be written as
\begin{eqnarray}
sq_{r}^{L}(s)+2q_{r}^{L}(s)-\frac{\mu}{d}\sum_{\delta}q_{r+\delta}^{L}(s) 
= \Big[b^{L}(s)+1\Big]\delta_{r,0},
\label{qL_eq}
\end{eqnarray}
where
\begin{eqnarray}
q_{r}^{L}(s)=\int_{0}^{\infty}dt\,e^{-st}q_{r}(t)
\label{laplace}
\end{eqnarray}
is the Laplace transform of $q_{r}(t)$ (similar formula connects 
$b^{L}(s)$ and $b(t)$).

The equation (\ref{qL_eq}) can be solved by introducing the Green function
\begin{eqnarray}
G_{r}^{L}(s,\mu)=\frac{1}{N}\sum_{p}\frac{e^{irp}}{s+2f_{\mu}(p)},
\label{green}
\end{eqnarray}
where $f$ is defined in (\ref{Gamma}), that satisfies
\begin{eqnarray}
sG_{r}^{L}(s,\mu)+2G_{r}^{L}(s,\mu)-\frac{\mu}{d}
\sum_{\delta}G_{r+\delta}^{L}(s,\mu) 
= \delta_{r,0}.
\label{green_eq}
\end{eqnarray}
Hence, the solution of the non-homogeneous differential equation 
(\ref{qL_eq}) is computed as
\begin{eqnarray}
q_{r}^{L}(s)=\sum_{r^{\prime}}G_{r-r^{\prime}}^{L}(s,\mu)
\Big[1+b^{L}(s)\Big]\delta_{r^{\prime},0}=\Big[1+b^{L}(s)\Big]
G_{r}^{L}(s,\mu).
\label{qs}
\end{eqnarray}
The function $b^{L}(s)$ is fixed remembering that the condition 
$q_{0}(t)=1$, or $q_{0}(s)=1/s$, should be satisfied. 
It is easy to see that
\begin{eqnarray}
b^{L}(s) = \frac{1}{sG_{0}^{L}(s,\mu)}-1,
\label{bs}
\end{eqnarray}
which implies
\begin{eqnarray}
q_{r}^{L}(s)=\frac{1}{s}\frac{G_{r}^{L}(s,\mu)}{G_{0}^{L}(s,\mu)}.
\label{qs_sol}
\end{eqnarray}
The stationary value for the pair correlation is obtained through 
the Laplace final value theorem
\begin{eqnarray}
q_{r}(\infty)=\lim_{t\to\infty}q_{r}(t)=\lim_{s\to 0}s\,q_{r}^{L}(s)
=\frac{G_{r}^{L}(0,\mu)}{G_{0}^{L}(0,\mu)}.
\label{qinf}
\end{eqnarray}

In the same fashion, the stationary value for $b(t)$ can be calculated as
\begin{eqnarray}
b(\infty)=\lim_{t\to\infty}b(t)=\lim_{s\to 0}s\,b^{L}(s)
=\frac{1}{G_{0}^{L}(0,\mu)}.
\label{binf}
\end{eqnarray}

\subsection{Susceptibility}
\label{susceptibility}

In this work, the susceptibility is defined through the (spatial) variance
\begin{eqnarray}
\nonumber \chi(t) & = & \sum_{r}\Big[ \langle
\delta(\sigma_{0},1)\delta(\sigma_{r},1)\rangle
-\langle\delta(\sigma_{0},1)\rangle\langle\delta(\sigma_{r},1)\rangle \Big] \\
 & = & \left(\frac{q-1}{q^{2}}\right)\sum_{r}q_{r}(t),
\label{susc}
\end{eqnarray}
where the random initial condition is being assumed.
Starting from a disordered state ($m_{j}(0)=0$ for any $j$) 
and assuming $\mu\neq 1$, the stationary susceptibility,
\begin{eqnarray}
\chi(\infty)=\lim_{t\to\infty}\chi(t) 
= \frac{q-1}{2q^{2}\left(1-\mu\right)}\frac{1}{G_{0}^{L}(0,\mu)},
\label{susc_inf}
\end{eqnarray}
is obtained by invoking the previous result (\ref{qinf}).
In a hypercubic lattice, one has
\begin{eqnarray}
G_{0}^{L}(0,\mu)\sim\left\{
\begin{array}{lcl}
\displaystyle\frac{1}{2}\left(\frac{d}{2\pi\mu}
\right)^{\frac{d}{2}}\Gamma\left(1-\frac{d}{2}\right)
\left(1-\mu\right)^{\frac{d-2}{2}}, &  & 0<d<2, \\
 & & \\
-\displaystyle\frac{1}{2\pi\mu}\ln\left(1-\mu\right), &  & d=2, \\
\end{array}
\right.
\label{greengamma}
\end{eqnarray}
as $\mu\sim 1$, and $\lim_{\mu\to 1}G_{0}^{L}(0,\mu)<\infty$ 
for $d>2$. Therefore, the stationary susceptibility is
\begin{eqnarray}
\chi(\infty)\sim\left\{
\begin{array}{lcl}
\displaystyle\left(\frac{q-1}{q^{2}}\right)
\left(\frac{2\pi}{d}\right)^{\frac{d}{2}}
\frac{1}{\Gamma\left(1-\frac{d}{2}\right)}
\left(1-\mu\right)^{-\frac{d}{2}}, &  & 0<d<2, \\
 & & \\
\pi\mu\displaystyle\left(\frac{q-1}{q^{2}}\right)
\frac{\left(1-\mu\right)^{-1}}
{\left[-\ln\left(1-\mu\right)\right]}, &  & d=2, \\
 & & \\
\left[\displaystyle\left(\frac{q-1}{2q^{2}}\right)
\frac{1}{G_{0}^{L}(0,\mu)}\right]\left(1-\mu\right)^{-1}, &  & d>2,
\end{array}
\right.\,
\label{chigamma}
\end{eqnarray}
from which the exponent $\gamma$ is obtained: the susceptibility diverges algebraically with 
exponent $d/2$ for $0<d<2$ and $1$ for $d\geq 2$ with 
logarithmic corrections for $d=2$.

\section{Two-time autocorrelation and response functions}
\label{CR}

\subsection{Two-time functions}
\label{CR1}

The analytical form for the autocorrelation and response function will 
be determined in this subsection in order to analyze the stationary and 
aging dynamical regimes. The two-time autocorrelation is defined as
\begin{eqnarray}
 C(t,t^{\prime}) = \lim_{N\to\infty}\frac{1}{N}\sum_{j}
\Big[\langle\delta(\sigma_{j}(t),1)\delta(\sigma_{j}(t^{\prime}),1)
\rangle-\langle\delta(\sigma_{j}(t),1)\rangle\langle
\delta(\sigma_{j}(t^{\prime}),1)\rangle\Big],
\label{C}
\end{eqnarray}
with the two-time correlation
\begin{eqnarray}
\langle\delta(\sigma_{j}(t),1)\delta(\sigma_{j}(t^{\prime}),1)\rangle 
= \sum_{\sigma}\sum_{\sigma^{\prime}}\delta(\sigma_{j}(t),1)
P(\sigma,t|\sigma^{\prime},t^{\prime})
\delta(\sigma_{j}^{\prime}(t^{\prime}),1)P(\sigma^{\prime},t^{\prime}),
\label{2-cor}
\end{eqnarray}
where $P(\sigma,t|\sigma^{\prime},t^{\prime})$ is the conditional 
probability of finding the configuration $\sigma$ at time $t$ given 
the configuration $\sigma^{\prime}$ at an earlier time $t^{\prime}$. 
Noting that
\begin{eqnarray}
\langle\delta(\sigma_{j}(t),1)\rangle=\sum_{\sigma}\delta(\sigma_{j}(t),1)
P(\sigma,t|\sigma^{\prime},t^{\prime})
\label{cond}
\end{eqnarray}
with the condition at time $t^\prime$ being 
$\langle\delta(\sigma_{j}(t^\prime),1)\rangle
=\delta(\sigma_{j}^{\prime}(t^{\prime}),1)$,
and invoking the definition (\ref{m}) and the equation (\ref{mt}), 
it is possible to show that
\begin{eqnarray}
C(t,t^{\prime}) = \lim_{N\rightarrow\infty}\left(\frac{q-1}{q^{2}}\right)\sum_{j}
\Gamma_{j}(t-t^{\prime})q_{j}(t^{\prime}),
\label{Ct}
\end{eqnarray}
for the disordered initial condition.

On the other hand, if one assumes an arbitrary initial condition, one has
\begin{eqnarray}
\nonumber C(t,t^{\prime}) & = & \lim_{N\rightarrow\infty}\Bigg[\left(\frac{q-1}{q^{2}}\right)\sum_{j}\Gamma_{j}(t-t^{\prime})q_{j}(t^{\prime})+\frac{\left(q-1\right)\left(q-2\right)}{q^{2}}\frac{1}{N}\sum_{j}m_{j}(t)- \\
 & & - \left(\frac{q-1}{q}\right)^{2}\frac{1}{N}\sum_{j}m_{j}(t)m_{j}(t^{\prime})\Bigg]\,.
\label{Ctm}
\end{eqnarray}

The evaluation of response function requires the presence of a (small) 
perturbation on the system. In the analysis of stochastic models, the 
introduction of an external field modifies the one-spin-flip rate to
\begin{eqnarray}
\nonumber w_{j}^{h}(\sigma) & = & w_{j}(\sigma)e^{h_{j}\delta(\sigma_{j},1)} \\
 & = & \left(\frac{1-\mu}{q}\right)\delta(\sigma_{j},1)
+\frac{h_{j}\mu}{2d}\sum_{\delta}\delta(\sigma_{j+\delta},1)\delta(\sigma_{1},1)
+\mathcal{O}(h_{j}^{2})\,,
\label{wh}
\end{eqnarray}
where a Taylor's expansion was performed in the last step.

Performing similar calculations of subsection \ref{site_m}, it is possible 
to show that
\begin{eqnarray}
\frac{dm_{j}(t)}{dt}=-m_{j}(t)+\frac{\mu}{2d}\sum_{\delta}m_{j+\delta}(t)
+\frac{1}{2q}b(t)h_{j}(t),
\label{dmdt_h}
\end{eqnarray}
assuming again disordered initial condition, when 
$m_{j}(t)=\mathcal{O}(h_{j})$. The solution of this differential 
equation, which can be obtained following the same previous ideas, is
\begin{eqnarray}
m_{j}(t) = \frac{1}{2q}\sum_{k}\int_{0}^{t}dt^{\prime}\,
\Gamma_{j-k}(t-t^{\prime})h_{k}(t^{\prime})b(t^{\prime}).
\label{m_h_sol}
\end{eqnarray}

The above result, (\ref{m_h_sol}), is sufficient to evaluate the 
two-time response function
\begin{eqnarray}
\nonumber R(t,t^{\prime}) & = & \lim_{N\to\infty}\frac{1}{N}
\sum_{j}\left.\frac{\delta\langle\delta(\sigma_{j}(t),1)
\rangle}{\delta h_{j}(t^{\prime})}\right|_{h\downarrow 0} \\
 & = & \left(\frac{q-1}{q}\right)\lim_{N\to\infty}\frac{1}{N}
\sum_{j}\left.\frac{\delta m_{j}(t)}{\delta h_{j}(t^{\prime})}
\right|_{h\downarrow 0} = \left(\frac{q-1}{2q^{2}}\right)
\Gamma_{0}(t-t^{\prime})b(t^{\prime}).
\label{R}
\end{eqnarray}

It is worth to stress that the formula (\ref{R}) for the autoresponse function is obtained even assuming an arbitrary initial condition.

\subsection{Stationary regime}
\label{stationary}

The stationary regime can be realized when both the waiting time 
($t^{\prime}$) and observational time ($t$) grow with the constraint that 
$\tau=t-t^{\prime}\geq 0$ is fixed. In this limit, and assuming disordered initial condition, the autocorrelation,
\begin{eqnarray}
C(t,t^{\prime})=C(\tau)=\frac{1}{G^{L}_{0}(0,\mu)}\int\frac{d^{d}p}{\left(2\pi\right)^{d}}\frac{e^{-f_{\mu}(p)\tau}}{2f_{\mu}(p)}\,,
\label{C_st}
\end{eqnarray}
and the response function,
\begin{eqnarray}
R(t,t^{\prime})=R(\tau)=\left(\frac{q-1}{2q^{2}}\right)
\frac{\Gamma_{0}(\tau)}{G_{0}^{L}(0,\mu)},
\label{R_st}
\end{eqnarray}
are functions of the time difference $\tau$ only, and they are 
related to the usual form of the fluctuation-dissipation relation 
$R(\tau)=\partial_{t^{\prime}}C(\tau)$, as expected in a stationary regime.

\subsection{Aging regime}
\label{aging}

The aging scenario can be seen when both the observational time ($t$) 
and waiting time ($t^{\prime}$) are made large without the difference 
$\tau=t-t^{\prime}$ being fixed. In the stationary regime, where the 
difference $\tau=t-t^{\prime}$ was fixed, the limit $t^{\prime}\to\infty$ 
made the function $q_{r}(t^{\prime})$ in (\ref{Ct}) and $b(t^{\prime})$ 
in (\ref{R}) time-independent. This is not the case in the aging regime, 
where both autocorrelation and response function depend on $t$ and 
$t^{\prime}$ independently. More precisely, the transient is observed 
if $t\gg t^{\prime}$, and this condition can be realized if the limit 
$t\to\infty$ is taken before the limit $t^{\prime}\to\infty$. Assuming disordered initial condition, and from previous results, it can be shown at criticality $\mu=1$ that in this regime the autocorrelation function behaves as
\begin{eqnarray}
C(t,t^{\prime})\sim\left\{
\begin{array}{lcl}
\displaystyle\left(\frac{q-1}{q^{2}}\right)\frac{2^{\frac{d}{2}+1}}{d}\frac{\sin\left(\frac{\pi d}{2}\right)}{\pi}\left(t-t^{\prime}\right)^{-\frac{d}{2}}t^{\prime^{\frac{d}{2}}} & , & 0<d<2 \\
 & & \\
2\displaystyle\left(\frac{q-1}{q^{2}}\right)\frac{t^{\prime}}{\left(t-t^{\prime}\right)\ln t^{\prime}} & , & d=2 \\
 & & \\
\displaystyle\left(\frac{q-1}{q^{2}}\right)\left(\frac{d}{2\pi}\right)^{\frac{d}{2}}\frac{1}{G_{0}^{L}(0,1)}\left(t-t^{\prime}\right)^{-\frac{d}{2}}t^{\prime} & , & d>2
\end{array}
\right.
\label{Cscaling}
\end{eqnarray}
and the response function is asymptotically equal to
\begin{eqnarray}
R(t,t^{\prime})\sim\left\{
\begin{array}{lcl}
\displaystyle\left(\frac{q-1}{2q^{2}}\right)2^{\frac{d}{2}}\frac{\sin\left(\frac{\pi d}{2}\right)}{\pi}\left(t-t^{\prime}\right)^{-\frac{d}{2}}t^{\prime^{\frac{d}{2}-1}} & , & 0<d<2 \\
 & & \\
2\displaystyle\left(\frac{q-1}{2q^{2}}\right)\frac{1}{\left(t-t^{\prime}\right)\ln t^{\prime}} & , & d=2 \\
 & & \\
\displaystyle\left(\frac{q-1}{2q^{2}}\right)\left(\frac{d}{2\pi}\right)^{\frac{d}{2}}\frac{1}{G_{0}^{L}(0,1)}\left(t-t^{\prime}\right)^{-\frac{d}{2}} & , & d>2
\end{array}
\right.\,.
\label{Rscaling}
\end{eqnarray}
The above results agree with the scaling $C(t,t^{\prime})\sim t^{\prime^{-b}}f_{C}(t/t^{\prime})$ and $R(t,t^{\prime})\sim t^{\prime^{-1-a}}f_{R}(t/t^{\prime})$\cite{HP07}, where $a=b=\left(d-2+\eta\right)/z$ (see table I), and $f_{C}$ and $f_{R}$ are scaling functions that behave as $f_{C/R}(t/t^{\prime})\sim A_{C/R}\left(t/t^{\prime}\right)^{-\lambda/z}$ for $t/t^{\prime}\sim\infty$.

In the aging regime, the fluctuation-dissipation theorem is not expected 
to hold anymore. The fluctuation-dissipation ratio
\begin{eqnarray}
X(t,t^{\prime})=\frac{R(t,t^{\prime})}
{\partial_{t^{\prime}}C(t,t^{\prime})},
\label{X}
\end{eqnarray}
which measures the distance of the model to the stationary state 
(when $X(t,t^{\prime})=1$), has the following limit:
\begin{eqnarray}
X(\infty,t^{\prime})=\lim_{t\to\infty}X(t,t^{\prime})
=\frac{b(t^{\prime})/2}{b(t^{\prime})-\left(1-\mu\right)\chi(t^{\prime})},
\label{X(inf,tprime)}
\end{eqnarray}
where $b(t)$ and $\chi(t)$ are given, respectively, by (\ref{binf}) 
and (\ref{susc_inf}). This result implies
\begin{eqnarray}
X_{\infty}=\lim_{t^{\prime}\to\infty}
\left[\lim_{t\to\infty}X(t,t^{\prime})\right]=\left\{
\begin{array}{lcl}
1,                        &  & \mu\neq 1, \\
 & & \\
\displaystyle\frac{1}{2}, &  & \mu=1,
\end{array}
\right.\,
\label{X_ag}
\end{eqnarray}
which is identical to the Ising case \cite{HSTdO06}.

The previous result has considered disordered initial condition, which assumes $m_{j}(t=0)=0$ for every site $j$. If one starts from an arbitrary initial condition, it is possible to show that (now using equations (\ref{Ctm}) and (\ref{R}))
\begin{eqnarray}
X({\infty},t^{\prime}) = \lim_{N\rightarrow\infty}\left[\frac{b(t^{\prime})/2}{b(t^{\prime})-\left(1-\mu\right)\sum_{j}q_{j}(t^{\prime})}\right]\,,
\end{eqnarray}
which shows that a similar formula for $X(\infty,t^{\prime})$ is obtained even for an arbitrary initial condition. The fluctuation-dissipation ratio $X_{\infty}$ is identical to (\ref{X_ag}); other non-trivial values for this ratio (for non-zero magnetization as initial condition) can be found, for instance, in \cite{GSPR05}, \cite{AS06} and \cite{CG07}.

\section{Dynamical exponents}
\label{dynamicalexponents}

\subsection{Dynamical exponent $\theta$}
\label{theta}

From the solution of equation (\ref{dmdt}),
\begin{eqnarray}
m_{j}(t)=m^{0}e^{-\left(1-\mu\right)t},
\label{mtheta}
\end{eqnarray}
one sees that at the critical point $\mu=1$ the magnetization is 
constant and does not vary with time. This implies the exponent 
$\theta$, defined through $m_{j}(t)\sim m^{0}t^{\theta}$ 
\cite{Z96}
in the short-time regime, to be zero. It is possible also to 
calculate this exponent by means of the time correlation of
the total magnetization \cite{TdO98}. 

\subsection{Dynamical exponent $\lambda/z$}
\label{lambdaz}

At the critical point $\mu=1$, one may calculate the dynamical 
exponents $\lambda$ and $z$, defined through 
$C(t,0)\sim t^{-\frac{\lambda}{z}}$. From (\ref{Ct}), it is immediate that
\begin{eqnarray}
C(t,0)=\left(\frac{q-1}{q^{2}}\right)\sum_{j}\Gamma_{j}(t)q_{j}(0).
\label{Ct0}
\end{eqnarray}
The $q_{r}(0)$ can be evaluated by invoking (\ref{qs_sol}) 
and the Laplace initial value theorem
\begin{eqnarray}
q_{r}(0)=\lim_{t^{\prime}\to 0^{+}}q_{r}(t^{\prime})
=\lim_{s\to\infty}sq_{r}^{L}(s)=\delta_{r,0}.
\label{initvalue}
\end{eqnarray}
In the thermodynamic limit, this result implies
\begin{eqnarray}
C(t,0)=\left(\frac{q-1}{q^{2}}\right)e^{-t}\left[I_{0}
\left(\frac{\mu}{d}t\right)\right]^{d}\sim
\left(\frac{q-1}{q^{2}}\right)\frac{e^{-\left(1-\mu\right)t}}
{\left(2\pi\mu/d\right)^{\frac{d}{2}}}t^{-\frac{d}{2}},
\label{Ct0_2}
\end{eqnarray}
where $I_{0}(x)$ is the modified Bessel function of order $0$ 
and the last passage is obtained in the asymptotic limit $t\gg 1$. 
The autocorrelation decays exponentially for $\mu\neq 1$; 
nevertheless, if $\mu=1$, one sees that
\begin{eqnarray}
\frac{\lambda}{z}=\frac{d}{2}.
\label{lambda/z}
\end{eqnarray}

\subsection{Dynamical exponent $\zeta$}
\label{zeta}

Another dynamical exponent of interest is $\zeta$, defined through 
$\chi(t)\sim t^{\zeta}$. From the results obtained in subsection \ref{pair}, 
the Laplace transform of the susceptibility can be written as
\begin{eqnarray}
\chi^{L}(s)=\left(\frac{q-1}{q^{2}}\right)\frac{1}{s^{2}G_{0}^{L}(s,\mu)}.
\label{Lchi}
\end{eqnarray}

The asymptotic behaviour of the (dynamical) susceptibility $\chi(t)$ 
for large times corresponds to the Laplace anti-transform of $\chi^{L}(s)$ 
when $s\sim 0$. In this regime, one can evaluate the Green function at 
criticality as
\begin{eqnarray}
G_{0}^{L}(s,1)\sim\left\{
\begin{array}{lcl}
\displaystyle\left(\frac{d}{4\pi}\right)^{\frac{d}{2}}
\Gamma\left(1-\frac{d}{2}\right)s^{\frac{d-2}{2}}, &  & 0<d<2, \\
 & & \\
-\displaystyle\frac{1}{2\pi}\ln s, &  & d=2, \\
 & & \\
G_{0}^{L}(0,1), &  & d>2,
\end{array}
\right.\,
\label{Gs0}
\end{eqnarray}
which yields
\begin{eqnarray}
\chi(t)\sim\left\{
\begin{array}{lcl}
\displaystyle\frac{2}{d}\left(\frac{q-1}{q^{2}}\right)\left(\frac{4\pi}{d}
\right)^{\frac{d}{2}}\frac{1}{\Gamma\left(\frac{d}{2}\right)
\Gamma\left(1-\frac{d}{2}\right)}t^{\frac{d}{2}}, &  & 0<d<2, \\
 & & \\
2\pi\displaystyle\left(\frac{q-1}{q^{2}}\right)\frac{t}{\ln t}, &  & d=2, \\
 & & \\
\left(\displaystyle\frac{q-1}{q^{2}}\right)\displaystyle\frac{1}
{G_{0}^{L}(0,1)}t, &  & d>2,
\end{array}
\right.\,
\label{dynchi}
\end{eqnarray}
showing that
\begin{eqnarray}
\zeta=\left\{
\begin{array}{lcl}
\displaystyle\frac{d}{2}, &  & 0<d<2, \\
 & & \\
1, &  & d\geq 2,
\end{array}
\right.\,
\label{expzeta}
\end{eqnarray}
with logarithmic corrections for $d=2$.

\subsection{Dynamical exponent $z$}
\label{z}

The exponent $z$, defined by the behaviour of the correlation length 
$\xi\sim\left(1-\mu\right)^{-\frac{\nu_{\parallel}}{z}}$ will be 
estimated through the spatial correlation, which can be casted as
\begin{eqnarray}
q_{r}(t=\infty)=\frac{G_{r}^{L}(0,\mu)}{G_{0}^{L}(0,\mu)}
=\left(\frac{d}{2\pi\mu}
\right)^{\frac{d}{2}}\left(\frac{\mu}{d}
\right)^{\frac{d-2}{4}}\frac{2^{\frac{d-2}{4}}
\left(1-\mu\right)^{\frac{d-2}{4}}}{G_{0}(0,\mu)r^{\frac{d-2}{2}}}
K_{\frac{d-2}{2}}\left(r\sqrt{\frac{2\left(1-\mu\right)d}{\mu}}\right),
\label{qr}
\end{eqnarray}
where $K_{\nu}(z)$ is the Macdonald's function, which behaves as
\begin{eqnarray}
K_{\nu}(z)\sim\left\{
\begin{array}{lcl}
\displaystyle\frac{2^{\nu-1}\Gamma(\nu)}{z^{\nu}}, & & 
|z|\ll 1\textrm{ and }\nu\neq 0, \\
 & & \\
\displaystyle\ln\left(\frac{2}{z}\right), &  & |z|\ll 1\textrm{ and }\nu=0, \\
 & & \\
\displaystyle\frac{e^{-z}}{\sqrt{2\pi z}}, &  & |z|\gg 1.
\end{array}
\right.\,
\label{macdonald}
\end{eqnarray}
Therefore, for large distances, the correlation decays exponentially 
as $e^{-r/\xi}$, where
\begin{eqnarray}
\xi = \sqrt{\frac{\mu}{2d}}\left(1-\mu\right)^{-\frac{1}{2}}
\label{correlationlength}
\end{eqnarray}
is associated to the correlation length. On the other hand, when the system 
is near criticality in the sense that $r\ll\xi$, one has
\begin{eqnarray}
q_{r}(t=\infty) & \sim &\left\{
\begin{array}{lcl}
\displaystyle\frac{\Gamma\left(\frac{d-2}{2}\right)}
{\Gamma\left(\frac{2-d}{2}\right)}\frac{1}{\left(r/2\xi\right)^{d-2}},
& & 0<d<2, \\
 & & \\
\displaystyle\frac{\ln\left(r/2\xi\right)}
{\ln\left(\sqrt{2\mu}/2\xi\right)},
&  & d=2, \\
 & & \\
\displaystyle\frac{d}{4\pi^{\frac{d}{2}}\mu}
\frac{\Gamma\left(\frac{d-2}{2}\right)}{G_{0}(0,\mu)}\frac{1}{r^{d-2}}, 
&  & d>2,
\end{array}
\right.\,
\label{qasymp}
\end{eqnarray}
from which it is possible to see that the exponent $\eta$\cite{B82} is equal to zero for $d>2$ and is equal to $2-d$ if $0<d\leq 2$. 
From (\ref{correlationlength}), the exponent $\nu_{\perp}$ (defined by 
$\xi\sim\left(1-\mu\right)^{-\nu_{\perp}}$) is equal to $1/2$, while 
the dynamical exponent $z=\nu_{\parallel}/\nu_{\perp}$ is then $z=2$ 
(since $\nu_{\parallel}=1$, as seen in subsection \ref{site_m}).

\begin{table}
\caption{Critical exponents for the nonequilibrium linear
$q$-state model}
\begin{center}
\begin{tabular}{|c|c|c|c|}\hline
\textrm{Exponent}  & $0<d\leq 2$ & $d>2$ \\\hline
 $\zeta$           & $d/2$       & $1$   \\\hline
 $z$               & $2$         & $2$   \\\hline
 $\lambda$         & $d$         & $d$   \\\hline
 $\theta$          & $0$         & $0$   \\\hline
\end{tabular}
\qquad\qquad
\begin{tabular}{|c|c|c|c|}\hline
\textrm{Exponent}  & $0<d\leq 2$ & $d>2$ \\\hline
 $\beta$           & $0$         & $0$   \\\hline
 $\nu_{\parallel}$ & $1$         & $1$   \\\hline
 $\nu_{\perp}$     & $1/2$       & $1/2$ \\\hline
 $\gamma$          & $d/2$       & $1$   \\\hline
 $\eta$            & $2-d$         & $0$   \\\hline
\end{tabular}
\end{center}
\label{exponents}
\end{table}

Collecting the previous results (see \ref{site_m}, 
\ref{susceptibility}, \ref{lambdaz}, \ref{zeta} and \ref{z}), 
the table \ref{exponents} is obtained.
One should recall that for $d>2$, one has $\zeta=1$ and $\gamma=1$ 
(the other exponents remain unchanged). These values for the exponents 
satisfy the relations $\zeta z=\left(d-2\beta/\nu\right)$ and 
$\theta z=\left(d-\lambda\right)$. It is worth mentioning that
these exponents are in agreement with the universality class 
of the voter model \cite{dOMS93,DCCH01}.

\section{Conclusions}
\label{conclusions}

This paper has established a number of exact calculations for the 
dynamical and static behaviour of the $d$-dimensional nonequilibrium
linear $q$-state lattice model, which is a generalization of the
nonequilibrium linear Glauber model.
This model is fully symmetric in the sense that it is invariant
under the permutation among all the $q$ states, having the
same symmetry of the equilibrium Potts model.
Although the analytical form of many functions are now 
distinct from the linear Glauber model, many similarities were reported in 
this paper. The stationary and aging regimes were both characterized,
with the usual 
fluctuation-dissipation relation satisfied in the former regime and violated 
in the later regime at the criticality $\mu=1$. The fluctuation-dissipation 
ratio $X_{\infty}$ indicates that the dynamical behaviour of the present 
model is similar to the linear Glauber model with Ising spins, which 
is just a particular case ($q=2$), and thus independent of
the number of states $q$.

When $\mu=1$, we recover the voter model with $q$-state and
in this case the system finds itself in the critical state.
From the results for the correlation functions it is possible
to generalize a result already known for the case $q=2$ 
about the stationary states.
In one and two dimensions the only possible stationary states
are the ones in which all the sites of the lattice are
in one of the $q$ absorbing states, which one depends on the initial 
configuration. This statements seems to be in contradiction with
the result (\ref{mtheta}) which says that the magnetization remains constant.
To understand this it suffices to remember the meaning of
choosing an initial condition with magnetization $m_0$. 
This means that one should consider several initial
configurations whose average gives the magnetization $m_0$.
Each one of these configuration will reach one of the $q$ absorbing states.
The averages over these absorbing states will give an average
$m$ which according to (\ref{mtheta}) should equal $m_0$.
In three or more dimensions there are other states
stationary states besides the $q$ absorbing states.

\section{Acknowledgements}
\label{acknowledgements}

We acknowledge the financial support of the project 
COFECUB-USP and INCT/CNPq de Fluidos Complexos. 
MOH is supported by the Brazilian agency CNPq.
We wish to acknowledge also helpful discussions with C. Chatelain.

\section{Appendix}

In this appendix (and also in the main text), the Landau notation was adopted:

\noindent
i) if $f(x)=\mathcal{O}(g(x))$ (assuming $g(x)>0$), then there exists 
$x_{0}$ such that $|f(x)|<Ag(x)$ for some constant $A$ if $x>x_{0}$.

\noindent
ii) if $f(x)=o(g(x))$ (assuming $g(x)>0$), then $\lim_{x\to\infty}
{f(x)}/{g(x)}=0$.

\subsection{Dynamical susceptibility (case $d=2$)}
\label{a_chi}

The asymptotic behaviour of the dynamical susceptibility is calculated 
through the Laplace anti-transform
\begin{eqnarray}
\chi(t)=\frac{1}{2\pi i}\int_{c-i\infty}^{c+i\infty}ds\,e^{st}\chi^{L}(s),
\label{a_chid2}
\end{eqnarray}
where $c$ is real and
larger than the real part of any pole of 
$\chi^{L}(s)$, given by (\ref{Lchi}). Since the $0<d<2$ and $d>2$ 
cases are simpler, the evaluation of dynamical susceptibility will be 
presented for $d=2$ only, which implies 
\begin{eqnarray}
\chi^{L}(s)\sim 2\pi
\left(\frac{q-1}{q^{2}}\right)\frac{1}{s^{2}\left(-\ln s\right)}.
\end{eqnarray}

One should first consider the integral
\begin{eqnarray}
B(t) = \frac{1}{2\pi i}\int_{c-i\infty}^{c+i\infty}ds\,\frac{e^{st}}
{s\left(-\ln s\right)},
\label{a_bromwitchd2}
\end{eqnarray}
where $c$ and $\tilde{c}$ are real and
larger than the real part of any 
pole of the integrand.
The function $B(t)$ relates to $\chi(t)$ through
\begin{eqnarray}
\frac{d}{dt}\chi(t)=2\pi\left(\frac{q-1}{q^{2}}\right)B(t).
\label{chiB}
\end{eqnarray}

\begin{figure}[tbp]
\centering
\epsfig{file = 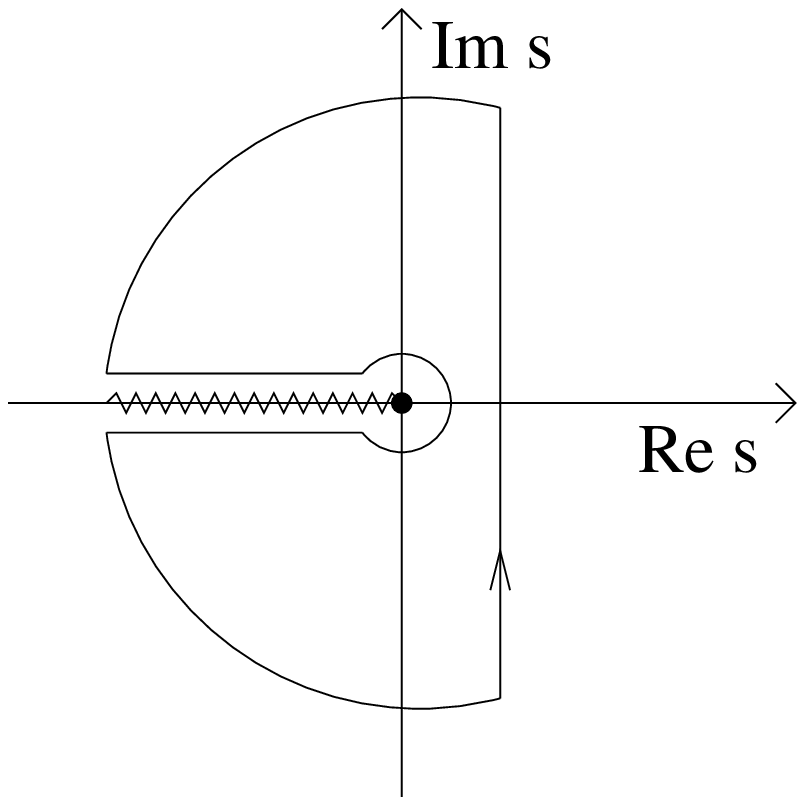, scale = 0.7}
\caption{Integration contour used in the evaluation of the function
  $B(t)$ given in (\ref{a_bromwitchd2}).}
\label{contour}
\end{figure}

By using the contour shown in figure \ref{contour}
and invoking the residue theorem, 
the integral (\ref{a_bromwitchd2}) can be casted in the form
\begin{eqnarray}
\nonumber B(t) & = & \int_{0}^{\infty}dr\,\frac{e^{-rt}}{r\left(\ln^{2}r
+\pi^{2}\right)} \\
 & = & B_{a}(t) + B_{b}(t) + B_{c}(t),
\label{a_B3}
\end{eqnarray}
where
\begin{eqnarray}
B_{a}(t) = \int_{0}^{\frac{1}{t\ln t}}dr\frac{e^{-rt}}{r\left(\ln^{2}r
+\pi^{2}\right)},
\label{a_ba}
\end{eqnarray}
\begin{eqnarray}
\quad B_{b}(t) = \int_{\frac{1}{t\ln t}}^{\frac{\ln t}{t}}dr
\frac{e^{-rt}}{r\left(\ln^{2}r+\pi^{2}\right)}
\label{a_bb}
\end{eqnarray}
and
\begin{eqnarray}
B_{c}(t) = \int_{\frac{\ln t}{t}}^{\infty}dr\frac{e^{-rt}}
{r\left(\ln^{2}r+\pi^{2}\right)}.
\label{a_bc}
\end{eqnarray}

These three functions, $B_{a}(t)$, $B_{b}(t)$ and $B_{c}(t)$ 
will be evaluated separately.

\bigskip
\noindent
\underline{Function $B_{a}(t)$}

\bigskip
Since
\begin{eqnarray}
B_{a}(t) = \int_{0}^{\frac{1}{t\ln t}}dr\frac{1}{r\left(\ln^{2}r
+\pi^{2}\right)}\left[1+\sum_{m=1}^{\infty}
\frac{\left(-rt\right)^{m}}{m!}\right],
\label{ba_1}
\end{eqnarray}
and
\begin{eqnarray}
\nonumber \left|\int_{0}^{\frac{1}{t\ln t}}dr\frac{1}
{r\left(\ln^{2}r+\pi^{2}\right)}\sum_{m=1}^{\infty}
\frac{\left(-rt\right)^{m}}{m!}\right| & \leq & 
\int_{0}^{\frac{1}{t\ln t}}dr\frac{1}{r\left(\ln^{2}r
+\pi^{2}\right)}\sum_{m=1}^{\infty}\left(rt\right)^{m} \\
\nonumber & \leq & \int_{0}^{\frac{1}{t\ln t}}dr\frac{1}
{r\ln^{2}r}\sum_{m=1}^{\infty}\left(\frac{1}{t\ln t}t\right)^{m} \\
\nonumber & \leq & \frac{1}{\ln t}\int_{0}^{\frac{1}{t\ln t}}dr
\frac{1}{r\ln^{2}r} \\
 & = & \frac{1}{\ln t}\frac{1}{\ln\left(t\ln t\right)} 
= \mathcal{O}\left(\frac{1}{\ln^{2}t}\right),
\label{ba_2}
\end{eqnarray}
then
\begin{eqnarray}
\nonumber B_{a}(t) & = & \int_{0}^{\frac{1}{t\ln t}}dr\frac{1}
{r\left(\ln^{2}r+\pi^{2}\right)}+\mathcal{O}\left(\frac{1}
{\ln^{2}t}\right) \\
\nonumber & = & \int_{-\infty}^{-\frac{1}{\pi}\ln\left(t\ln t\right)}\frac{dy}
{\pi\left(y^{2}+1\right)}+\mathcal{O}\left(\frac{1}{\ln^{2}t}\right) \\
 & = & -\frac{1}{\pi}\left\{\arctan\left[\frac{1}{\pi}
\ln\left(t\ln t\right)\right]-\frac{\pi}{2}\right\}
+\mathcal{O}\left(\frac{1}{\ln^{2}t}\right),
\label{ba_3}
\end{eqnarray}
where the change of variable $r\to e^{\pi y}$ was performed 
in the second line. For $t\gg 1$, one has
\begin{eqnarray}
B_{a}(t) = \frac{1}{\ln\left(t\ln t\right)}+\mathcal{O}
\left(\frac{1}{\ln^{2}t}\right) = \frac{1}{\ln t}
+\mathcal{O}\left(\frac{\ln\left(\ln t\right)}{\ln^{2}t}\right).
\label{ba_4}
\end{eqnarray}

\bigskip
\noindent
\underline{Function $B_{b}(t)$}

\bigskip
The function $B_{b}(t)$ has the following upper bound:
\begin{eqnarray}
\nonumber \left|B_{b}(t)\right| & = & \left|
\int_{\frac{1}{t\ln t}}^{\frac{\ln t}{t}}dr\frac{e^{-rt}}
{r\left(\ln^{2}r+\pi^{2}\right)}\right|\leq\frac{1}
{\ln^{2}t}\int_{\frac{1}{t\ln t}}^{\frac{\ln t}{t}}dr\frac{e^{-rt}}{r}
=\frac{1}{\ln^{2}t}\int_{\frac{1}{\ln t}}^{\ln t}du\frac{e^{-u}}{u} \\
 & \leq & \frac{e^{-\frac{1}{\ln t}}}{\ln^{2}t}\int_{\frac{1}{\ln t}}^{\ln t}
\frac{du}{u}=\frac{2e^{-\frac{1}{\ln t}}\ln\left(\ln t\right)}{\ln^{2}t}
=\mathcal{O}\left(\frac{\ln\left(\ln t\right)}{\ln^{2}t}\right).
\label{bb_1}
\end{eqnarray}

\bigskip
\noindent
\underline{Function $B_{c}(t)$}

\bigskip
The function $B_{c}(t)$ has the following upper bound:
\begin{eqnarray}
\nonumber \left|B_{c}(t)\right| & = & \left|
\int_{\frac{\ln t}{t}}^{\infty}dr\frac{e^{-rt}}{r\left(\ln^{2}r
+\pi^{2}\right)}\right| \leq \frac{1}{\pi^{2}}
\int_{\frac{\ln t}{t}}^{\infty}dr\frac{e^{-rt}}{r} \leq 
\frac{1}{\pi^{2}}\frac{1}{\frac{\ln t}{t}}\int_{\frac{\ln t}{t}}^{\infty}dr 
e^{-rt} \\
 & = & \frac{1}{\pi^{2}t\ln t} = \mathcal{O}\left(\frac{1}{t\ln t}\right).
\label{bc_1}
\end{eqnarray}

From (\ref{a_B3}), (\ref{ba_4}), (\ref{bb_1}) and (\ref{bc_1}), 
one finally has
\begin{eqnarray}
B(t)=\frac{1}{\ln t}+\mathcal{O}\left(\frac{\ln\left(\ln t\right)}
{\ln^{2}t}\right),
\label{a_Bend}
\end{eqnarray}
which can be inserted in (\ref{chiB}) to yield
\begin{eqnarray}
\nonumber \chi(t) & \sim & 2\pi\left(\frac{q-1}{q^{2}}\right)\int dt B(t) 
= 2\pi\left(\frac{q-1}{q^{2}}\right)\left[\int dt\left(\frac{1}{\ln t}
-\frac{1}{\ln^{2}t}\right) + \int dt\frac{1}{\ln^{2}t}\right] \\
 & \sim & 2\pi\left(\frac{q-1}{q^{2}}\right)\frac{t}{\ln t}\Big[1+o(1)\Big].
\label{chid2}
\end{eqnarray}

\subsection{Spatial correlation function}
\label{a_qr}

In the thermodynamic limit ($N\to\infty$), the spatial correlation 
function can be casted as
\begin{eqnarray}
q_{r}(t\to\infty)=\frac{G_{r}(0,\mu)}{G_{0}(0,\mu)}
=\frac{1}{G_{0}(0,\mu)}
\int_{[-\pi,\pi)^{d}}\frac{d^{d}p}{\left(2\pi\right)^{d}}
\frac{e^{i\vec r\cdot\vec p}}{\epsilon+2\mu\left(1-\frac{1}{d}
\sum_{i=1}^{d}\cos p_{i}\right)},
\label{a_qr_1}
\end{eqnarray}
as seen in subsection \ref{pair}, and $\epsilon=2\left(1-\mu\right)$. 
It is straightforward to see also that
\begin{eqnarray}
\nonumber q_{r}(t\to\infty) & 
= & \frac{1}{G_{0}(0,\mu)}\int_{[-\pi,\pi)^{d}}
\frac{d^{d}p}{\left(2\pi\right)^{d}}\int_{0}^{\infty}du 
e^{-u\left[\epsilon+2\mu\left(1-\frac{1}{d}
\sum_{i=1}^{d}\cos p_{i}\right)\right]}e^{i\vec r\cdot\vec p} \\
 & = & \frac{1}{G_{0}(0,\mu)}\int_{0}^{\infty}\frac{dy}
{\epsilon}e^{-y-\frac{2\mu y}{\epsilon}}\prod_{i=1}^{d}
\int_{-\pi}^{\pi}I_{r_{i}}\left(\frac{2\mu y}{\epsilon d}\right),
\label{a_qr_2}
\end{eqnarray}
where $I_{\nu}(z)$ is the modified Bessel function, which 
behaves asymptotically as (\cite{SP85})
\begin{eqnarray}
I_{\nu}(z\sim\infty)\sim\frac{e^{z-\frac{\nu^{2}}{2z}}}{\sqrt{2\pi z}}.
\label{a_qr_3}
\end{eqnarray}
Therefore, when the system approaches the critical point 
($\epsilon\sim 0^{+}$), one has
\begin{eqnarray}
\nonumber q_{r}(t\to\infty) & = & \frac{1}{G_{0}(0,\mu)}
\left(\frac{d}{4\pi\mu}\right)^{\frac{d}{2}}
\epsilon^{\frac{d-2}{2}}\int_{0}^{\infty}dy\,e^{-y-\frac{r^{2}\epsilon d}
{4\mu y}} \\
 & = & \left(\frac{d}{2\pi\mu}\right)^{\frac{d}{2}}\frac{1}{G_{0}(0,\mu)}
\frac{\left[2\left(1-\mu\right)\right]^{\frac{d-2}{2}}}
{\left(r/\xi\right)^{\frac{d-2}{2}}}K_{\frac{d-2}{2}}\left(r/\xi\right),
\label{a_qr_4}
\end{eqnarray}
where $K_{\nu}(z)$ is the Macdonald's function (or modified Bessel 
function of the third kind)
\begin{eqnarray}
K_{\nu}(z) = \frac{1}{2}\left(\frac{z}{2}\right)^{\nu}
\int_{0}^{\infty}dy\,e^{-y-\frac{z^{2}}{4y}}t^{-\nu-1},
\quad\qquad |\arg z|<\frac{\pi}{4},
\label{a_qr_5}
\end{eqnarray}
and
\begin{eqnarray}
\xi = \sqrt{\frac{\mu}{2d}}\left(1-\mu\right)^{-\frac{1}{2}}.
\label{a_qr_6}
\end{eqnarray}


\end{document}